\documentclass[aps,prl,twocolumn,showpacs,superscriptaddress]{revtex4}
\usepackage{graphicx}

\newcommand{\lam}{\ensuremath{\Lambda^0} }
\newcommand{\lamnospace}{\ensuremath{\Lambda^0}}
\newcommand{\lams}{\ensuremath{\Lambda^0\mathrm{s}} }

\newcommand{\sigz}{\ensuremath{\Sigma^0} }
\newcommand{\sigstar}{\ensuremath{\Sigma^*} }

\newcommand{\cascade}{\ensuremath{\Xi} }

\newcommand{\pim}{\ensuremath{\pi^-} }

\newcommand{\ra}{\ensuremath{\rightarrow} }
\newcommand{\dll}{\ensuremath{D_{LL'}^\Lambda} }

\newcommand{\dllnospace}{\ensuremath{D_{LL'}^\Lambda}}

\newcommand{\dllf}{\ensuremath{D_{LL',f}^\Lambda} }

\newcommand{\ubar}{\ensuremath{\overline{u}} }

\newcommand{\sbar}{\ensuremath{\overline{s}} }

\newcommand{\qlam}{\ensuremath{q_f^\Lambda} }

\newcommand{\dqlam}{\ensuremath{\Delta q_f^\Lambda} }

\begin{document}

\title{
  Longitudinal Spin Transfer to the $\Lambda$ Hyperon
 in Semi-Inclusive Deep-Inelastic Scattering 
}

% List of Institute Addresses

\def\groupalberta{\affiliation{Department of Physics, University of Alberta, Edmonton, Alberta T6G 2J1, Canada}}
\def\groupargonne{\affiliation{Physics Division, Argonne National Laboratory, Argonne, Illinois 60439-4843, USA}}
\def\groupbari{\affiliation{Istituto Nazionale di Fisica Nucleare, Sezione di Bari, 70124 Bari, Italy}}
\def\groupbeijing{\affiliation{School of Physics, Peking University, Beijing 100871, China}}
\def\groupchina{\affiliation{Department of Modern Physics, University of Science and Technology of China, Hefei, Anhui 230026, China}}
\def\groupcolorado{\affiliation{Nuclear Physics Laboratory, University of Colorado, Boulder, Colorado 80309-0390, USA}}
\def\groupdesy{\affiliation{DESY, 22603 Hamburg, Germany}}
\def\groupzeuthen{\affiliation{DESY, 15738 Zeuthen, Germany}}
\def\groupdubna{\affiliation{Joint Institute for Nuclear Research, 141980 Dubna, Russia}}
\def\grouperlangen{\affiliation{Physikalisches Institut, Universit\"at Erlangen-N\"urnberg, 91058 Erlangen, Germany}}
\def\groupferrara{\affiliation{Istituto Nazionale di Fisica Nucleare, Sezione di Ferrara and Dipartimento di Fisica, Universit\`a di Ferrara, 44100 Ferrara, Italy}}
\def\groupfrascati{\affiliation{Istituto Nazionale di Fisica Nucleare, Laboratori Nazionali di Frascati, 00044 Frascati, Italy}}
\def\groupgent{\affiliation{Department of Subatomic and Radiation Physics, University of Gent, 9000 Gent, Belgium}}
\def\groupgiessen{\affiliation{Physikalisches Institut, Universit\"at Gie{\ss}en, 35392 Gie{\ss}en, Germany}}
\def\groupglasgow{\affiliation{Department of Physics and Astronomy, University of Glasgow, Glasgow G12 8QQ, United Kingdom}}
\def\groupillinois{\affiliation{Department of Physics, University of Illinois, Urbana, Illinois 61801-3080, USA}}
\def\groupmichigan{\affiliation{Randall Laboratory of Physics, University of Michigan, Ann Arbor, Michigan 48109-1040, USA }}
\def\groupmoscow{\affiliation{Lebedev Physical Institute, 117924 Moscow, Russia}}
\def\groupnikhef{\affiliation{Nationaal Instituut voor Kernfysica en Hoge-Energiefysica (NIKHEF), 1009 DB Amsterdam, The Netherlands}}
\def\groupstpetersburg{\affiliation{Petersburg Nuclear Physics Institute, St. Petersburg, Gatchina, 188350 Russia}}
\def\groupprotvino{\affiliation{Institute for High Energy Physics, Protvino, Moscow region, 142281 Russia}}
\def\groupregensburg{\affiliation{Institut f\"ur Theoretische Physik, Universit\"at Regensburg, 93040 Regensburg, Germany}}
\def\grouprome{\affiliation{Istituto Nazionale di Fisica Nucleare, Sezione Roma 1, Gruppo Sanit\`a and Physics Laboratory, Istituto Superiore di Sanit\`a, 00161 Roma, Italy}}
\def\groupsimonfraser{\affiliation{Department of Physics, Simon Fraser
 University, Burnaby, British Columbia V5A 1S6, Canada}}
 \def\groupjefferson{\affiliation{Thomas Jefferson National
 Accelerator Facility, Newport News, Virginia 23606, USA}}
\def\grouptriumf{\affiliation{TRIUMF, Vancouver, British Columbia V6T 2A3, Canada}}
\def\grouptokyo{\affiliation{Department of Physics, Tokyo Institute of Technology, Tokyo 152, Japan}}
\def\groupamsterdam{\affiliation{Department of Physics and Astronomy, Vrije Universiteit, 1081 HV Amsterdam, The Netherlands}}
\def\groupwarsaw{\affiliation{Andrzej Soltan Institute for Nuclear Studies, 00-689 Warsaw, Poland}}
\def\groupyerevan{\affiliation{Yerevan Physics Institute, 375036 Yerevan, Armenia}}
\def\groupnone{\noaffiliation}

% Set Institute Order

\groupalberta
\groupargonne
\groupbari
\groupbeijing
\groupchina
\groupcolorado
\groupdesy
\groupzeuthen
\groupdubna
\grouperlangen
\groupferrara
\groupfrascati
\groupgent
\groupgiessen
\groupglasgow
\groupillinois
\groupmichigan
\groupmoscow
\groupnikhef
\groupstpetersburg
\groupprotvino
\groupregensburg
\groupjefferson
\grouprome
\groupsimonfraser
\grouptriumf
\grouptokyo
\groupamsterdam
\groupwarsaw
\groupyerevan

% List of Authors

\author{A.~Airapetian}  \groupmichigan
\author{N.~Akopov}  \groupyerevan
\author{Z.~Akopov}  \groupyerevan
\author{M.~Amarian}  \groupzeuthen \groupyerevan
\author{A.~Andrus}  \groupillinois
\author{E.C.~Aschenauer}  \groupzeuthen
\author{W.~Augustyniak}  \groupwarsaw
\author{R.~Avakian}  \groupyerevan
\author{A.~Avetissian}  \groupyerevan
\author{E.~Avetissian}  \groupfrascati
% ...

\collaboration{The HERMES Collaboration} \noaffiliation

\date{\today}

\begin{abstract}
The transfer of polarization from a high-energy positron  to a
\lam hyperon produced in semi-inclusive deep-inelastic scattering
has been measured. The data have been obtained by the HERMES experiment at
DESY using the 27.6 GeV
longitudinally  polarized positron beam of the HERA
 collider and unpolarized gas targets internal  to the
positron (electron)  storage ring.
The longitudinal spin transfer coefficient is found
to be $\dll = 0.11 \pm 0.10\ \mathrm{(stat)} \pm 0.03\ \mathrm{(syst)}$
at an average fractional energy carried by the \lam hyperon $\langle z
\rangle= 0.45$.
The dependence of \dll on both
the fractional energy $z$ and the fractional longitudinal momentum $x_F$
is  presented.

\end{abstract}

% insert suggested PACS numbers in braces
\pacs{13.88+e,13.60.-r,13.60.Rj}

\maketitle
%==========================================================================
% Introduction
%==========================================================================

In this paper  the study of the longitudinal spin transfer from a
polarized positron  to a \lam hyperon produced in the
deep-inelastic scattering  process is presented.
 The measurements are   sensitive
to two unknowns: the spin structure of the lightest hyperon,
and the spin-dependent dynamics of the fragmentation process in
deep-inelastic scattering.

Given the non-trivial spin structure of the proton
~\cite{SpinReview}, it is of interest to consider the spin
structure of other baryons. In this respect the \lam hyperon is
particularly interesting, as it is the lightest strange baryon of
the SU(3) spin-$\frac{1}{2}$ octet based on up ($u$), down ($d$),
and strange ($s$) quarks. The number density 
for quarks plus antiquarks of flavor $f$
in the \lam hyperon is denoted below as  $q^\Lambda_f$ ($f=u,d,s$).

In the naive Constituent Quark Model the spin of the \lam hyperon is
entirely carried by the $s$ quark:
${\Delta q^\Lambda_s  = 1}$,
 while  the $ud$ pair is in a spinless
(singlet) state, i.e.,
${\Delta q^\Lambda_u=\Delta q^\Lambda_d=0}$. Here ${\Delta
q^\Lambda_f \equiv  q^{\Lambda+}_f -  q^{\Lambda-}_f}$, where
$q^{\Lambda+}_f$ and $q^{\Lambda-}_f$ describe the net alignment of
the quark  spins along ($+$) or against ($-$) the hyperon spin
direction, respectively, while  the unpolarized number density is
 ${ q^\Lambda_f \equiv q^{\Lambda+}_f +  q^{\Lambda-}_f}$.
 Alternatively, one can use SU(3)-flavor
symmetry in conjunction with the experimental results on the proton
to estimate the first moments of the
helicity-dependent quark distributions in the \lam hyperon. Using such
assumptions
Burkardt and Jaffe  found
$\Delta q^\Lambda_u = \Delta q^\Lambda_d =-0.23 \pm 0.06$
and
$\Delta q^\Lambda_s=0.58 \pm 0.07$~\cite{BJ}.
According to this estimate, the spins of
the $u$ and $d$ quarks and antiquarks
are  directed predominantly opposite to the spin of the \lam
hyperon  resulting in    a weak but non-zero net polarization.
If such an SU(3)-flavor rotation (see Eq.~3 of Ref.~\cite{QCDSF}, for
example) is applied to the recent semi-inclusive
data on the nucleon~\cite{HERMES_semi}, the values
$\Delta q^\Lambda_u = \Delta q^\Lambda_d = -0.09 \pm 0.06$ and
$\Delta q^\Lambda_s = 0.47 \pm 0.07$
are obtained instead, favoring a much smaller polarization of the
$u$ and $d$ quarks and antiquarks. 
A  lattice-QCD calculation~\cite{QCDSF} also
finds small light-quark polarizations,
$\Delta q^\Lambda_u = \Delta q^\Lambda_d = -0.02 \pm 0.04$ and
$\Delta q^\Lambda_s = 0.68 \pm 0.04$.
Finally, other
authors~\cite{Ma00,Boros,Liang} have employed phenomenological models
to explore the dependence of $\dqlam(x)$ on the Bjorken scaling variable
$x$. These models predict a large positive polarization of the
$u$ and $d$ quarks in the kinematic region $x > 0.3$.

As it is not experimentally feasible to scatter directly from hyperon
targets, another probe of hyperon spin structure must be found to address
these model predictions.
One possibility, as suggested in Ref. \cite{Jaffe}, is to study hyperons
produced in the \textit{final state}
of the deep-inelastic scattering (DIS) process, and to employ 
the fragmentation process by which they
are formed as a ``polarimeter'' for the quarks within.
More precisely, when a longitudinally-polarized lepton beam is
scattered at high energies from a nucleon target,
angular momentum conservation dictates that  quarks of a particular
spin orientation  participate predominantly in the interaction. The
outgoing struck
quark is thus polarized, and hyperons produced  from its fragmentation
may ``remember'' its spin orientation and carry a longitudinal polarization
themselves.
Formally such a correlation may be expressed in terms of a
spin-dependent fragmentation function, denoted $G_{1,f}^{\Lambda}(z)$ in
the notation of Ref.~\cite{Bible},
where $z$  is the fractional energy of the \lam hyperon.
(This fragmentation function
has often appeared in the literature with different symbols, most notably
as  $\Delta D^\Lambda(z)$  in Ref.~\cite{Ma00}
or as $\Delta \hat{q}_\Lambda(z)$  in Ref.~\cite{Jaffe}.)
The magnitude of this spin-dependent fragmentation function
is sensitive to quark helicity conservation and to the correlation between
quark spins in the complex fragmentation process. It is  also
sensitive  to the spin structure of the produced  \lam hyperon,
provided the amount of \lamnospace s produced from unpolarized quarks
in the process of color-string breaking 
or from the decay of heavier hyperons with different spin structure
is not significant.

The polarization of final-state \lam hyperons can be measured via the
weak decay channel $\Lambda^0 \rightarrow p \pi^-$
through the angular distribution of the final-state particles:
\begin{equation}
  \frac{dN}{d\Omega_p}
  \propto 1 + \alpha \vec{P}_{\Lambda} \cdot \hat{k}_p.
  \label{eq:decay}
\end{equation}
Here $\frac{dN}{d\Omega_p}$ is the angular distribution of
the protons,
$\alpha = 0.642 \pm 0.013$ is the asymmetry parameter of the
parity-violating weak decay, $\vec{P}_{\Lambda}$ is the
polarization of the \lamnospace, and $\hat{k}_p$ is the unit
vector along the proton momentum in the rest frame of the
\lamnospace. Because of the parity-violating nature of this decay,
the proton is preferentially emitted along the spin direction of
its parent, thus offering  access to  spin degrees of freedom in
the deep-inelastic scattering final state.

Longitudinal spin transfer to \lam hyperons has previously been explored
by the LEP
experiments OPAL and ALEPH at an energy corresponding to the $Z^0$
pole~\cite{OPAL,ALEPH}.
In these experiments the \lam hyperons are predominantly
produced via the decay $Z^0 \ra s\sbar$, in which the primary strange
quarks from the decay are strongly (and negatively) polarized at the level
of $-91$\%. The OPAL and ALEPH data show a \lam polarization of about
$-30$\%
at $z>0.3$. It  rises  in magnitude as $z$ increases.
Here, $z$ is the ratio between the energy of the \lam and that of the
primary   (fragmenting) quark.
The LEP  data have been described  using a Lund-based Monte Carlo model
along with the following hypotheses, originally postulated in
Ref.~\cite{Gustafson}: (1) that the primary quarks
produced in the $Z^0$ decay retain their helicity throughout the
fragmentation process, (2) that the quarks produced from color-string
breaking have no preferred spin direction, and (3) that the spin structure
of the produced hyperons can be adequately described by the
Constituent Quark Model.

In contrast to the LEP experiments, production of \lam hyperons in
deep-inelastic scattering of leptons from nucleons is dominated by scattering
from $u$ and $d$ quarks.
In the NOMAD
experiment~\cite{NOMAD}, the production of \lam hyperons was studied in
$\nu_\mu$ charged-current interactions. Also in contrast to the LEP
experiments, the NOMAD data are concentrated in the kinematic domain $x_F < 0$
corresponding to the so-called target fragmentation region.
(The Feynman-$x$ variable is defined in the
standard way as $x_F \equiv p_\parallel/p_{\parallel max}$. Here
$p_\parallel$ is the projection of the hadron momentum on
the virtual-photon ($\gamma^* $) or $W$-boson ($W^*$) direction,
$p_{\parallel max}$
is its maximum possible value,
and all quantities
are evaluated in the $\gamma^* N$ or $W^* N$ center-of-mass system,
where $N$ is the target nucleon).
A non-zero longitudinal \lam polarization was observed at $x_F<0$
while at $x_F>0$ the polarization
was found to be  compatible with zero ~\cite{NOMAD}.
A mechanism  giving rise to non-zero polarization values
in the region   $x_F < 0$ is described in Ref.~\cite{Ellis}.

Using charged lepton beams, only two measurements of longitudinal \lam
polarization in deep-inelastic scattering have been reported to date.
The E665 collaboration~\cite{E665}
measured a negative polarization by using a polarized muon beam of 470 GeV.
The statistical accuracy of the E665 experiment is
rather limited, however, as only 750 \lam events were identified.

Results on the longitudinal spin transfer
in deep-inelastic scattering  were also reported by the
HERMES collaboration~\cite{HERMES_DLL}. These data were obtained using
the 27.6 GeV polarized positron beam of the HERA $e$-$p$ collider,
and were collected during the years 1996 and 1997.
After subtraction of the background and application of several
kinematic requirements, about 2,000 \lam hyperons were reconstructed.
The longitudinal spin-transfer coefficient,
which is defined below in  Eq.~\ref{eq:depol},
for forward ($x_F > 0$) \lam
production  was measured to be
$\dll = 0.11 \pm 0.17\ \mathrm{(stat)} \pm 0.03\ \mathrm{(sys)}$, at an
average fractional  energy $\langle z \rangle= 0.45$.
In deep-inelastic scattering, the fractional  energy $z$ is defined as
$z =E_{\Lambda}/\nu$, where $E_{\Lambda}$ is the energy of the
\lam hyperon, $\nu=E-E'$, and
$E$ and $E'$ represent the energy of the primary and
scattered lepton, respectively. Since the energy of the current (struck)
quark after  absorption of the virtual photon
is very close to  $\nu$, the value  $z$ in deep-inelastic scattering is
practically the same as that in the LEP experiments discussed above,
thus allowing comparison of both results.

Recently, the CLAS experiment at Jefferson Laboratory measured
large spin transfers from  polarized 2.6 GeV  beam
electrons  to  \lams produced in the exclusive reaction
$\vec{e} p \rightarrow e' K^+ \vec{\Lambda}$~\cite{CLAS}.
However, due to the exclusive
nature of this reaction and the low energy of the experiment, these data
cannot be readily compared with results from high-energy deep-inelastic
scattering.

The HERMES results presented in this paper surpass the data of
Ref.~\cite{HERMES_DLL} in  statistical  precision.
The new data    were  mostly accumulated
during the very successful HERA data-taking period in the years 1999-2000.
In this period  a Ring Imaging \v{C}erenkov
(RICH) detector ~\cite{HERMES_RICH}
was used for hadron identification.
The old HERMES \dll data collected during
the years 1996 and 1997 (Ref.~\cite{HERMES_DLL}) are included in the
analysis, leading to a total of almost 8,000 \lam events.

As compared to previous measurements, the additional  data allow  the
exploration of
the $z$-dependence of the spin transfer \dllnospace.
This is of particular interest, as the $z$-dependence provides a
crucial test for the dominant \lamnospace-formation mechanism. Some models
actually predict a very pronounced $z$-dependence for $\Lambda^0$
polarization. The data were also binned in the variable $x_F$,
enabling a comparison  of all the  available data  collected in
the target- and current-fragmentation  regions
by the HERMES, NOMAD, and E665   experiments.

The paper is organized as follows.
Section 2 summarizes the
spin transfer in the framework of the quark-parton model.
Having introduced the relevant variables in section 2, section 3 is devoted
to a brief description of the experiment and a discussion of the analysis
techniques. The experimental results are presented and discussed
in section 4, and section 5 concludes the paper with summary remarks.

%==========================================================================
% QPM Framework: Longitudinal Spin Transfer
%==========================================================================

\section{II. Longitudinal Spin Transfer}
\label{section:DLL}

\begin{figure}[ht]
  \begin{center}
  \includegraphics[width=\columnwidth]{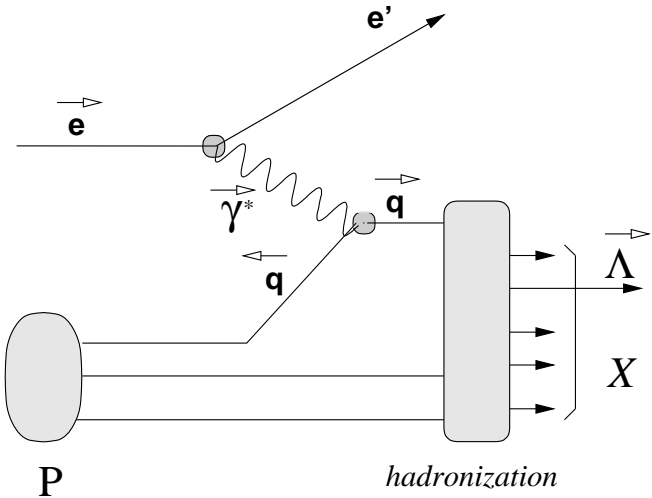}
  \caption{
     The single-quark scattering mechanism leading to \lam
     production in  polarized deep-inelastic positron scattering.
  }
  \label{fig:diagram}
  \end{center}
\end{figure}

The dominant mechanism for semi-inclusive production of longitudinally
polarized \lam hyperons in polarized deep-inelastic scattering is sketched
in Fig.~\ref{fig:diagram}.
It should be noted that  at the  moderate energy
of the HERMES experiment the distinction between the current and target
fragmentation domain  is not very sharp. The condition  $x_F>0$ only
selects
\lam particles  moving forward in the $\gamma^*p$ rest frame but does not
exclude  a contribution from the target remnant.
Nevertheless, with this requirement imposed
the remnant  contribution  is assumed to be  reduced.

As indicated by the arrows in
Fig.~\ref{fig:diagram},
a beam positron  of   positive helicity emits a polarized
virtual photon (denoted $\gamma^*$) which is
absorbed by a quark $q$
in the target proton with the spin direction opposite to that of
$\vec{\gamma^*}$.   This fixes the
spin orientation of the struck quark: after the  spin-1 photon
is absorbed,
the outgoing quark has the same  helicity  as   the virtual
photon.

%..........................................................................
% Figure 1: kinematic diagram
%..........................................................................

More precisely, if the longitudinal polarization of the beam is given by $P_b$
and the target is unpolarized, the struck quark will acquire a polarization
$P_q = P_b D(y)$
directed along its momentum. Here $y = \nu/E$ is the fractional energy
carried
by the photon  and $D(y) \simeq [ 1-(1-y)^2 ] / [1 + (1-y)^2 ]$ is the
depolarization factor taking into account the loss  of polarization of the
virtual photon as compared   to that  of the incident positron.
Positive beam polarization $P_b$ refers to the case when the beam
positron has preferentially positive helicity in the target rest
frame.
The component of the polarization transferred along the direction
$L'$ from the virtual photon to the produced \lam is given by
\begin{equation}
  P^\Lambda_{L'} = P_b D(y) \dllnospace,
  \label{eq:depol}
\end{equation}
where $L$ is the primary quantization axis, directed along
the virtual photon momentum.
The spin transfer coefficient \dll  in Eq.~\ref{eq:depol}
describes the probability that the polarization of the struck quark
is transferred to the \lam hyperon along the secondary
quantization axis $L'$.
In principle, the spin transfer can be studied experimentally 
for any direction of the $L'$ axis. Interesting information on
the dynamics of the reaction and on the mechanism of spin transfer
from the struck quark to the produced \lam can be obtained by
measuring the longitudinal component of the transferred polarization.
Different authors differ in their definition of the direction of
the longitudinal component~\cite{Jaffe,Bible,Kotz98}: choices for
the $L'$ direction include the directions of the momentum of the 
virtual-photon, of the produced \lamnospace, 
and of the lepton beam. As described in later sections, 
two of these choices were explored in this analysis
and gave fully-compatible results.

In the notation of~\cite{Bible}, the longitudinal spin-transfer
from the virtual photon to the \lam hyperon in DIS is expressed in terms
of a spin-transfer fragmentation function $G_{1,f}$:
\begin{equation}
\label{eq:dll}
  \dll(x,z,Q^2) = \frac
    { \sum_f e_f^2 q_f(x,Q^2) G_{1,f}^\Lambda(z,Q^2) }
    { \sum_f e_f^2 q_f(x,Q^2) D_{1,f}^\Lambda(z,Q^2) }.
\end{equation}

Here,  $e_f$
is the charge of the quark (or antiquark),
and the sum is taken over quark  (antiquark) flavors $f$.
The function $q_f(x,Q^2)$ is the number density of a quark
$f$ in the target, and   $x=Q^2/2M_p\nu$ represents the  Bjorken scaling
variable,  where $M_p$ is the proton mass  and  $Q^2$ is the negative
four-momentum transfer squared.

In Eq.~\ref{eq:dll},
$D_{1,f}^\Lambda(z,Q^2)$ is the familiar spin-independent
fragmentation function describing the number density for \lam
production from a primary quark $f$. Less familiar is the
spin-dependent fragmentation function
$G_{1,f}^\Lambda$~\cite{Bible}. It is defined as
$G_{1,f}^\Lambda=D_{1,f+}^{\Lambda+} - D_{1,f+}^{\Lambda-}$, while
the unpolarized fragmentation function is
$D_{1,f}^\Lambda=D_{1,f+}^{\Lambda+} + D_{1,f+}^{\Lambda-}$. Here, the
symbols $D_{1,f+}^{\Lambda+}$ or  $D_{1,f+}^{\Lambda-}$  are used
to  denote the fragmentation functions for a quark of helicity $+$
to produce a \lam of helicity $+$ or $-$,  respectively. It is
assumed that  $D_{1,f+}^{\Lambda+}=D_{1,f-}^{\Lambda-}$ and
$D_{1,f+}^{\Lambda-}=D_{1,f-}^{\Lambda+}$.

Both the quark-density distributions and the fragmentation  functions in
Eq.~\ref{eq:dll}
are slowly varying with $Q^2$, so that to a good approximation
\[\dll(x,z,Q^2)\simeq \dll(x,z)_{Q^2=\langle Q^2\rangle},\]
where $\langle Q^2 \rangle$ is the  average value of $Q^2$. 
After integrating over $x$ it reads:
\begin{eqnarray}
\label{eq:dllpar}
 \dll(z)&\simeq &\sum_f  \frac
{G_{1,f}^\Lambda(z)}{D_{1,f}^\Lambda(z)} \nonumber
\int \frac
        { e_f^2 q_f(x) D_{1,f}^\Lambda(z)}
{ \sum_{f'} e_{f'}^2 q_{f'}(x)
D_{1,{f'}}^\Lambda(z) }dx \nonumber\\
 &=& \sum_f \dllf(z)\; \omega_f^\Lambda(z).
\end{eqnarray}
Here  the quantity
\dllf  denotes the   {\it partial} spin transfer from a
struck quark of flavor $f$ to a \lam hyperon:
\begin{equation}
  \dllf (z) =
 \frac{ G_{1,f}^\Lambda (z) }{ D_{1,f}^\Lambda (z) }
        \equiv
 \frac{ D_{1,f+}^{\Lambda+} (z) - D_{1,f+}^{\Lambda-} (z) }
        { D_{1,f+}^{\Lambda+} (z) + D_{1,f+}^{\Lambda-} (z) }.
  \label{eq:dllq}
\end{equation}

The purity $\omega_f^\Lambda (z)$   in Eq.~\ref{eq:dllpar} is the
net probability that a \lam was produced at average 
$Q^2\approx \langle Q^2 \rangle$ with a  fractional energy $z$
after  absorption of a virtual photon by a quark of flavor $f$.
It is obvious that
\begin{equation}
 \sum_f \omega_f^\Lambda (z)=1.
\label{purnorm}
\end{equation}
%..........................................................................
% Figure X: purities
\begin{figure}[ht]
  \begin{center}
  \includegraphics[width=\columnwidth]{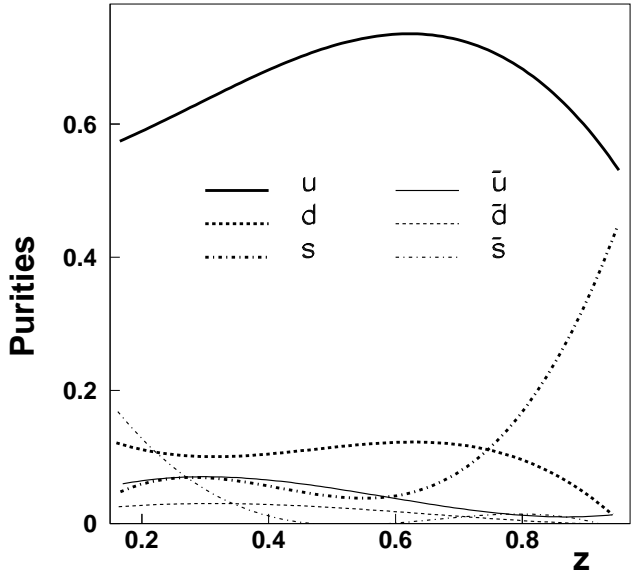}
  \caption{
        Purities for \lam production from the proton target within the
	HERMES acceptance, calculated separately for quarks 
	and antiquarks of various flavors
	at $x_F>0$ with $\langle Q^2 \rangle $=2.41 GeV$^2$.
  }
  \label{fig:subproc_purities}
  \end{center}
\end{figure}
%..........................................................................

The purities  depend on unpolarized quantities only, and can  be
obtained from  a   Monte Carlo model.
Fig.~\ref{fig:subproc_purities} shows
purity distributions for quarks of various flavors
calculated using the JETSET Monte Carlo for \lam production
from a proton target. The calculations have been done  in the current
fragmentation region ($x_F>0$) for HERMES kinematics.
The strength of the electromagnetic interaction between a lepton and a
quark  is proportional to the square of the quark charge $e_f$.
Hence \lam production
in electron (or muon) induced deep-inelastic scattering
is  dominated by scattering from $u$ quarks, as  shown in
Fig.~\ref{fig:subproc_purities}.
The strange quark plays a
minor role at moderate $z$. Its contribution is sharply increasing only at
very high $z$, which is difficult to access experimentally.

It is
apparent that the spin-transfer fragmentation function $G_{1,f}^\Lambda$
is expected to be related
to the spin structure of the \lamnospace \hspace{0.1cm} hyperon.
For example, under  the assumptions
that the produced hyperon actually contains the struck quark of flavor $f$,
that it was produced directly from fragmentation (and not from the decay
of a heavier hyperon), and
that the original helicity of the quark is preserved during the fragmentation
process, the partial spin-transfer coefficient has been estimated using
a theoretical model of the \lam  spin structure to be~\cite{Ashery}
\begin{equation}
  \dllf
    \equiv \frac{ G_{1,f}^\Lambda }{ D_{1,f}^\Lambda }
    \simeq \frac{\dqlam}{\qlam}.
  \label{eq:dll_deltaq}
\end{equation}
Here, $\dqlam/\qlam$ may be interpreted as the average polarization of
quarks of flavor $f$ in
the \lam hyperon,  and $\dllf$ as the
partial spin transfer averaged over  production kinematics.
Despite the number of simplifying assumptions that have  been
made in deriving  Eq.~\ref{eq:dll_deltaq}, it is a useful starting point
for developing a qualitative understanding of \dllnospace.

It must be noted  that  Eq.~\ref{eq:dll_deltaq}  can be also obtained as  a
consequence of the reciprocity relation
based on crossing symmetry~\cite{Grib}. Strictly speaking, however, it is
expected to be valid only at large values of the Bjorken scaling
variable for the \lam, $x_\Lambda$, and of $z$, providing an
exact link between the spin-transfer coefficient and  spin
structure of the \lam baryon  in the limit of $x_\Lambda
\rightarrow 1$, $z \rightarrow 1$.

Because of strong $u$-quark dominance,
one would expect that for electron (or muon) deep-inelastic scattering
\begin{equation}
  \dll \approx \Delta q^\Lambda_u/q^\Lambda_u.
  \label{eq:dll_u}
\end{equation}
This relation is not changed by the $d$-quark
contribution to the extent that
${\Delta q^\Lambda_u/q^\Lambda_u=\Delta q^\Lambda_d/q^\Lambda_d}$
because of isospin symmetry.
Further, as shown in Fig.~\ref{fig:subproc_purities}, the 
$u/d$-quark dominance approximation in Eq.~\ref{eq:dll_u} is exact for 
at least 70\% of the events within the accessible $z$-range, reaching 
almost 90\% at intermediate $z$ values of around 0.6.
In the  Constituent Quark Model,
the net $u$-quark polarization in the \lam is zero.
As explained in the introduction, the use of  recent HERMES
results on  proton structure gives a small negative  $\Delta
q^\Lambda_u$ of $-0.09 \pm 0.06$~\cite{HERMES_semi}, while a
lattice QCD calculation gives $-0.02 \pm 0.04$~\cite{QCDSF}. The
spin transfer to \lam hyperons in deep-inelastic scattering  might
thus be, in principle, a probe of the small non-strange
components of the \lam spin structure. This is quite different from the
case of $e^+ e^- \ra \lam X$, where the $s$-quark plays a dominant role.

%==========================================================================
% Experiment
%==========================================================================

\section{III. Experiment and Event Selection}
\label{section:experiment}

The \lam electroproduction data presented in this paper were
accumulated by the HERMES experiment at DESY. In this experiment,
the 27.6 GeV longitudinally-polarized positron  beam ~\cite{HERA}
of the HERA $e$-$p$ collider is passed
through an open-ended tubular storage cell into which polarized or
unpolarized target atoms in undiluted gaseous form are continuously
injected. The HERMES detector is described in detail in
Ref.~\cite{HERMES_spectro}.

The data presented here were recorded during  two two-year periods:
1996-1997 and 1999-2000 using positron beams.
A variety of unpolarized  target gases were used in the analysis.
Most of the data were collected from hydrogen and deuterium, but
$^{3}$He, $^{4}$He, $^{14}$N, $^{20}$Ne and $^{84}$Kr
targets were also included, and the data from all targets were combined.

The scattered  positrons  and the \lam decay products were detected by the
HERMES spectrometer
in the polar-angle range from 40 to 220 mrad.
A positron  trigger was formed from a coincidence between three
scintillator
hodoscope planes and a lead-glass calorimeter. The trigger required
a minimum energy deposit in the calorimeter of 3.5 GeV for the
data employed in this analysis.
Charged-particle identification was
based on the responses of four detectors; a threshold \v{C}erenkov counter, a
transition-radiation detector, a preshower scintillator hodoscope, and a
lead-glass calorimeter. Altogether, the particle identification (PID) system
provides an average positron identification efficiency of 99\% with a
hadron contamination of less than 1\%. In 1998 the threshold \v{C}erenkov
counter was replaced by the  Ring Imaging \v{C}erenkov
(RICH) detector~\cite{HERMES_RICH},
providing an improved hadron identification capability.

The \lam hyperons were identified in the analysis through their
$p \pim$ decay channel. Events were selected by requiring
the presence of at least three reconstructed tracks: a positron track and
two hadron candidates of opposite charge. If more than one positive or
negative hadron was found in one event, all possible combinations of
positive and negative hadrons were used.  The requirements
$Q^2 > 0.8$ GeV$^2$ and $W > 2$ GeV   were imposed
on the positron kinematics to ensure that the events
originated from the  deep-inelastic scattering domain.
Here $W = \sqrt{M_p^2 + 2 M_p \nu - Q^2}$ is the invariant
mass of the photon-nucleon system.
In addition, the requirement $y=1-E'/E < 0.85$ was imposed as the minimum
value of $E'$ was given  by the calorimeter threshold of 3.5 ~GeV.

The kinematics of the \lam decay products detected by the HERMES
spectrometer is such that the proton momentum is always much higher
than that of the pion.
These low-momentum pions are often bent so
severely in the spectrometer magnet that they fail to reach the
tracking chambers and PID detectors in the backward half of the spectrometer.
However, it is possible to evaluate the momentum of such ``short tracks''
using the hits recorded by the HERMES Magnet Chambers, a series of
proportional chambers located between the poles of the
spectrometer magnet~\cite{MagCh}.
The acceptance for \lam hyperons can be increased by almost a factor
of two by including in the analysis
the decay pions detected as short tracks.  As the great
majority of low-momentum particles produced in deep-inelastic scattering
are pions, particle identification is not essential for these tracks.
By comparison, particle identification of the decay proton
is important for background reduction, and PID is crucial in the
identification of the scattered  lepton. Candidates for these particles
were therefore required to be ``long tracks'',i.e., tracks  passed through
all detectors of the spectrometer.

\begin{figure}[ht]
  \begin{center}
  \includegraphics[width=\columnwidth]{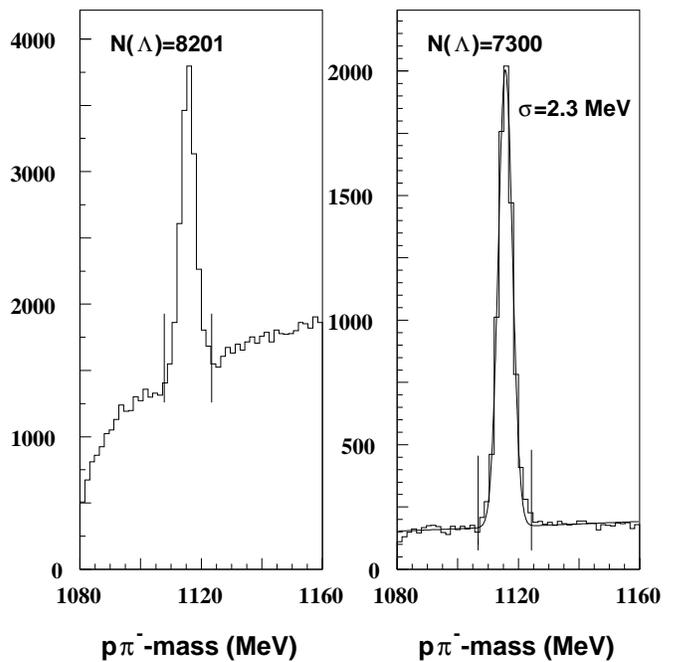}
  \caption{
    The yield of semi-inclusively produced \lam hyperons in
    deep-inelastic scattering. The left (right) panel shows the
    invariant-mass spectrum before (after) the application of
    background suppression cuts. The vertical lines show
    the boundaries at $\pm 3.3\,\sigma$.
    The spectra include essentially
    all data recorded from unpolarized targets by the HERMES spectrometer
    in the years 1996 -- 1997 and 1999 -- 2000, corresponding to a yield of
    $30.3 \times 10^6$ inclusive deep-inelastic scattering events.
  }
  \label{fig:invmass}
  \end{center}
\end{figure}

Two spatial vertices were reconstructed for each event by determining
the intersection (i.e., point of closest approach) of pairs of
reconstructed
tracks. The primary (production) vertex was determined from the intersection
of the beamline and the scattered beam lepton, while the secondary (decay)
vertex was determined from the intersection of the proton and pion tracks.
In both cases, the distance of closest approach was required to be less
than 1.5 cm.
All tracks were also required to satisfy a series of fiducial-volume cuts
designed to avoid the inactive edges of the detector.
For tracks fulfilling these requirements the invariant mass of
the hadron pair was evaluated, under the
assumption that the high-momentum leading hadron is
the proton while  the low-momentum hadron is the pion. There is
a clear \lam peak even
without background suppression cuts. This spectrum is displayed in
Fig.~\ref{fig:invmass}, left panel.

%..........................................................................
% Figure 2: invariant mass
%..........................................................................

In order to suppress background, two different approaches were taken,
depending on the availability of the RICH detector.
For the data taken  prior to
1998, hadron pairs with leading pions were suppressed  with the help of
the  threshold \v{C}erenkov counter. In addition, hadrons
emitted from the primary vertex were suppressed by introducing a vertex
separation requirement of $z_2 - z_1 > 10$~cm,
with $z_1$ and $z_2$ representing
the coordinates of the primary and secondary vertex positions along the
beam direction. In later years, with the RICH detector available, no
vertex separation cut was used.
In this case, protons with momenta larger
than 4~GeV (which applies to 75\% of all protons from \lam decay) could be
distinguished from lighter hadrons, providing sufficient background
reduction. Without background suppression, 8,200 \lam
events were extracted from all unpolarized data,
while the final data sample,
with all requirements imposed, contained 7,300 \lam events.
These numbers were obtained by integrating the \lam peak between
the boundaries $\pm 3.3\,\sigma$, shown in Fig.~\ref{fig:invmass}
with the vertical lines, and subtracting the  background.

An average beam polarization of about 55\% was typical during data taking.
Reversal of the polarization direction was performed three times during
the 1996-1997 data taking period, but more frequently thereafter.
Similar amounts of data
were recorded in each beam helicity state.

Two independent polarimeters were used to measure the beam polarization.
They used  similar techniques based on laser Compton
backscattering. The polarimeter TPOL~\cite{HERMES_TPOL}
measured the transverse beam polarization
outside the HERMES spin rotators, while the  polarimeter
LPOL~\cite{HERMES_LPOL}, located between the spin
rotators, measured the longitudinal polarization  at the HERMES
interaction point.
In practice, the value measured by the LPOL was taken, except for periods
where only the TPOL was in operation. The fractional systematic
uncertainty of  the beam polarization measured by the LPOL was
typically less than 2\%. For the TPOL  measurements,
it was less than 3.5\%.

%==========================================================================
% Experimental Results
%==========================================================================

\section{IV. Experimental Results}
\label{section:results}

%--------------------------------------------------------------------------
% Experimental Results: extraction procedure
%--------------------------------------------------------------------------
\subsection{A. Extraction of  \dll}
Combining Eqs.  \ref{eq:decay} and \ref{eq:depol}, the angular
distribution of decay protons may be expressed  in terms of the
longitudinal spin-transfer coefficient \dllnospace:
\begin{equation}
  \frac{ dN }{ d\Omega_{p} } \propto
          1 + \alpha P_b D(y) \dll \cos\Theta_{pL'}.
  \label{eq:decay_dll}
\end{equation}
Here $\Theta_{pL'}$ is the angle between the proton
momentum  in the \lam rest frame and the \lam spin quantization axis $L'$.
Proceeding from the discussion  in section~II,
two choices of the spin-quantization axis
$L'$  for the final-state \lam are  considered  in this analysis:
\begin{description}
  \item
    Axis~1: along the direction of the virtual photon momentum in the
    \lam rest frame;
  \item
    Axis~2: along the  direction of \lam momentum (not affected by the
    relativistic transformation to the \lam rest frame).
\end{description}
For HERMES kinematics, unlike the case of deep-inelastic scattering
at very high
energies,  axes~1 and 2    are  typically  not collinear.
It has been found  that  the angle   between  axis~1 and axis~2
varies over  a   wide range  with an average  value of 35~degrees
in the \lam rest frame.
Hence, both possibilities of \dll reconstruction
(axis~1 and axis~2) were considered in the analysis.

The HERMES spectrometer is a forward detector with a limited
acceptance for the reconstruction of \lam hyperons.
The  efficiency to detect  a pion from \lam decay  depends strongly
on its momentum in the  laboratory  frame, and  on its decay
angle ($\pi-\Theta_{pL'}$) in the \lam rest frame, resulting in
a  forward ($\cos\Theta_{pL'}>0$)/ backward
($\cos\Theta_{pL'}<0$) asymmetric  acceptance function.

In order to cancel  this  acceptance effect, the spin transfer to
the \lam has been determined by combining the two data sets
measured with opposite beam polarizations  into one helicity-balanced 
data sample, in which  the luminosity-weighted average
beam polarization for the selected data   is
\begin{equation}
  \overline{P_b} \equiv \frac{1}{L} \int P_b\,dL= 0.
\end{equation}
Here $L = \int dL$ is the integrated  luminosity.

A detailed derivation based on the method of maximum likelihood
leads to the    relation~\cite{SBform98}:
\begin{equation}
  D_{L L'}^\Lambda =
    \frac{1}{\alpha \overline{P_b^2}} \cdot
    \frac{\sum_{i=1}^{N_\Lambda} P_{b,i}\,D(y_i) \cos\Theta^i_{pL'}}
         {\sum_{i=1}^{N_\Lambda} D^2(y_i) \cos^2\Theta^i_{pL'}}.
  \label{eq:extract}
\end{equation}
Here, $\overline{P_b^2} \equiv (\frac{1}{L}) \int P_b^2\,dL$
is the luminosity-weighted average of the square of the beam polarization.

As follows from Eq.11, $D_{LL'}$ can be extracted from the data on
      an event-by-event basis using experimentally-measured values
      only.
As the beam polarization was reversed every six weeks,
 no  acceptance corrections are needed.
       On the other hand, in order to experimentally
      estimate the possible level of false asymmetries and related
      systematic uncertainties, a process with similar event topology
      where $D_{LL'}$ is necessarily equal to zero should be analyzed.
      The best candidate is $K^0_s$-meson production, with its  
subsequent
      weak decay to $\pi^+ \pi^-$: as the $K^0_s$ is a spin-zero meson,
      it cannot be polarized.

The \lam polarization was   studied
as a function of certain kinematic variables.
Using Eq.~\ref{eq:extract}, the spin transfer
${\hat D}^{\Lambda}_{LL'}$
was calculated for the   events within the \lam
invariant-mass peak (between the boundaries  $\pm 3.3\,\sigma$)
in each kinematic bin. The  fraction of  background events
$\epsilon=\frac{N_{bgr}}{N_\Lambda+N_{bgr}}$
within  the peak was typically of order  20\%.
The spin transfer  for the background $D^\Lambda_{LL'bgr}$     was
determined from  the events  above and below the peak   outside of
the $\pm 3.3\,\sigma$ invariant-mass window. In order to obtain
the final result for the net \lam events, the  spin transfer
within the \lam peak was corrected   for this   background
contribution in each kinematic bin as
\begin{equation}
  \label{Lambgr}
  \dll=\frac{{\hat D}^{\Lambda}_{LL'}-\epsilon D^\Lambda_{LL'bgr}}
  {1-\epsilon}.
\end{equation}
It should be noted that the  sideband (background)
spin-transfer coefficient
$D^\Lambda_{LL'bgr}$ in Eq.~\ref{Lambgr} was  always consistent
with zero.

%--------------------------------------------------------------------------
% Experimental Results: kinematic averages
%--------------------------------------------------------------------------
\subsection{B. \dll averaged over kinematics}

Table~\ref{table:results_ave1D} presents  the results for
\dll averaged over \mbox{HERMES} kinematics
with the requirement  $x_F > 0$  imposed.
As shown in the table, the results
from the 1996-1997 data set (where the RICH detector was not present)
are fully compatible with those from the 1999-2000 data set.
Hereafter, only results from the combined data set are  considered.

%..........................................................................
% Table 1: results averaged over kinematics, 1D extraction
\begin{table}
  \begin{center}
  \begin{tabular}{c||c|c|c}
     & 96-97 & 99-00 & All Data \\
    \hline
    \hline
    $\langle \dll \rangle$,
    axis 1 & $0.12 \pm 0.17$ & $0.12 \pm 0.12$ & $0.12 \pm 0.10$ \\
    $\langle \dll \rangle$,
    axis 2 & $0.13 \pm 0.17$ & $0.10 \pm 0.13$ & $0.11 \pm 0.10$ \\
    \hline
     $\int L$, pb$^{-1}$&226.9 &330.5&557.4\\
     $N_\Lambda$ & 2,452 & 4,294 & 6,746  \\
     $\langle z \rangle$ & 0.44 & 0.46 & 0.45 \\
     $\langle x_F \rangle$ & 0.29 & 0.31 & 0.30 \\
    \hline
  \end{tabular}
  \caption{Results for \dll
     averaged over kinematics with the  requirement
    $x_F > 0$ imposed.
    (Note that the total number of \lam events $N_\Lambda$ is reduced
    as compared with   that in Fig.~\ref{fig:invmass}, right panel,
    due to this requirement.) The quoted uncertainties are
    statistical only. The systematic uncertainties are  on the level
    of $\pm 0.03$ as  discussed in the text.
  }
  \label{table:results_ave1D}
  \end{center}
\end{table}
%..........................................................................

Three sources of systematic uncertainties were identified
and evaluated. First, the helicity-balanced analysis method
outlined above relies on an accurate normalization of the data samples with
positive and negative beam helicity.  In order to
estimate the associated systematic
uncertainty, the luminosity of each sample was determined using two
different methods: (1) using the number of inclusive
deep-inelastic scattering events found in   each
sample, and (2) using the number of semi-inclusive
deep-inelastic scattering  events containing   an
oppositely-charged hadron pair with invariant mass outside the \lam peak.
In both cases the spectrometer acceptance is assumed to be unaffected
by the reversal of the beam polarization. The inclusive
deep-inelastic scattering cross section
is independent of  the beam polarization  as the target is unpolarized.
The same is true for any semi-inclusive cross-section as long as it is
fully
integrated over the angular distribution of the final-state hadrons.
The spin-transfer results were found to differ by less than 0.03 when
the two normalization methods were used,
and a systematic uncertainty of $\pm 0.015$ was assigned
to account for this difference.

Second,  as a further check of the systematic uncertainties involved in
the extraction
procedure, the spin transfer coefficient was determined for
oppositely-charged hadron pairs ($h^+ h^-$), where the identity of the
hadrons was not restricted.
The invariant mass of each pair was calculated assuming it was a $p \pi^-$
pair with a mass lying
%    outside
%the \lam mass window shown in Fig.~\ref{fig:invmass}, left panel.
outside the \lam  window in the mass range indicated
      in Fig.~\ref{fig:invmass}.
 No background suppression cuts were applied in this
      case (Fig.~\ref{fig:invmass}, left panel). 
In semi-inclusive hadron production,
      hadron kinematics may be sensitive to the sign of the beam
      polarization \cite{Bible,Bacchetta:2002ux,Bacchetta:2003vn}.
      This may, in principle, result in correlations between
      $cos\Theta_p$ and $P_b$, and thus, in nonzero values of \dll
      for $h^+h^-$ pairs. These correlations, however, vanish provided
      that the target nucleon is unpolarized and the cross section is
      fully integrated over the angular distribution of the final-state
      hadrons. A statistically significant nonzero value of \dll for
      $h^+h^-$ pairs, though a priori not evident, would most likely be
      an indication of an instability of experimental conditions over
      the relatively long time of the data taking. Further, theoretical
      concerns aside, the measured spin-transfer to $h^+h^-$ pairs outside
      the \lam mass window provides a conservative estimate of
      the systematic error on \dll.
Using  this background   sample, \dll was found to be
compatible with zero:
$0.005 \pm 0.014$ using axis 1
and
$0.003 \pm 0.014$ using axis 2.
Nevertheless, a $\pm 0.014$ contribution was added to the overall 
systematic uncertainty to account for the statistical limitations of this
false-asymmetry test.

Third, semi-inclusive $K^0_s$ production has been studied in the
       experiment using the same kinematic cuts as those applied to
       the $\Lambda$ data set. 
        For the $K^0_s$ sample of 14800 events,
       the spin transfer coefficient was found to be compatible with
       zero: $D_{LL'}^{K^0_s} = 0.005 \pm 0.08$.

The systematic uncertainty of \dll due to beam polarization measurements
was estimated to be less than  0.002. Other uncertainties related to
smearing effects, choice of the background-suppression procedure, and
corrections for the background contribution (Eq.~\ref{Lambgr}) were also found
to be  small.

Based on these  results,  one can conclude  that the  systematic
uncertainty on the  spin-transfer coefficient is dominated by the
normalization uncertainty of the helicity-balanced analysis method
($\pm0.015$) and  possible experimental false asymmetries,
estimated with the help  of hadron pairs ($\pm 0.014$).
The overall systematic uncertainty of  the measured spin
transfer is thus estimated to be $\pm 0.03$.

As the measured value for \dll shows no significant dependence
on the choice of the longitudinal spin-quantization axis,
the results of this section can be summarized by a single value:
\begin{equation}
  \dll = 0.11 \pm 0.10\ \mathrm{(stat)} \pm 0.03\ \mathrm{(syst)}.
  \label{eq:result_dllave}
\end{equation}
This represents the  spin transfer to the \lam along its
momentum direction, averaged over the kinematic region
with $Q^2>0.8$ GeV$^2$, $y<0.85$, and $x_F>0$.
The average fractional energy of the \lam hyperons in this sample is
$\langle z \rangle = 0.45$,
the average momentum transfer  $\langle Q^2\rangle$=2.4 GeV$^2$,
and the average Bjorken variable $\langle x\rangle$=0.088.

%--------------------------------------------------------------------------
% Experimental Results: kinematic dependences
%--------------------------------------------------------------------------
\subsection{C. Dependence of \dll on $z$ and $x_F$}

\begin{figure}[ht]
  \begin{center}
    \includegraphics[width=\columnwidth]{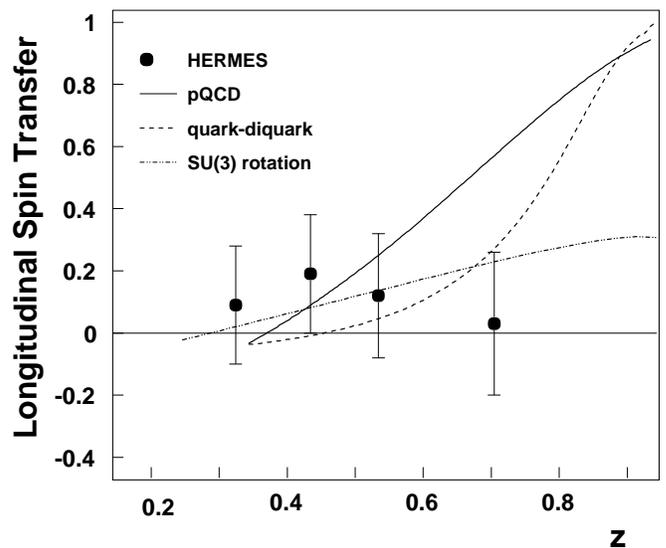}
    \caption{
      Dependence of the longitudinal spin-transfer coefficient \dll
      on $z$, for $x_F > 0$.
      The curves represent the phenomenological model calculations
      of Refs.~\protect\cite{Ma00,Ma02}, as described in the text.
      Error bars are statistical only.
    }
    \label{fig:dll_z}
  \end{center}
\end{figure}

The dependence of \dll on the energy fraction $z$
with the requirement  of $x_F > 0$  imposed
is presented in Fig.~\ref{fig:dll_z} and Table~\ref{table:dll_z}.
As the values measured in all bins are consistent for the two
axis choices, only the  results for axis ~1  are displayed in
Fig.~\ref{fig:dll_z}. Superimposed on the data are the
phenomenological model calculations of Ref.~\cite{Ma00} (pQCD and
quark-diquark models) which predict a pronounced rise of the spin
transfer at high values of $z$, and those of the model of
Ref.~\cite{Ma02} (SU(3)-flavor rotation of proton values) which
predicts a more gradual increase. Although the data presented here
extend to the highest values of $z$ yet explored in deep-inelastic
scattering, they display no evidence of a strong kinematic
dependence. One should remember, however,  that in the theoretical
models discussed above, the  \lam hyperon is assumed to be 
produced directly from the polarized struck quark, i.e., the
contribution from the heavier hyperon decays is not included in these
models (see  subsection D).

%..........................................................................
% Figure 4: z-dependence of DLL + theory curves
\begin{figure}[ht]
  \begin{center}
    \includegraphics[width=\columnwidth]{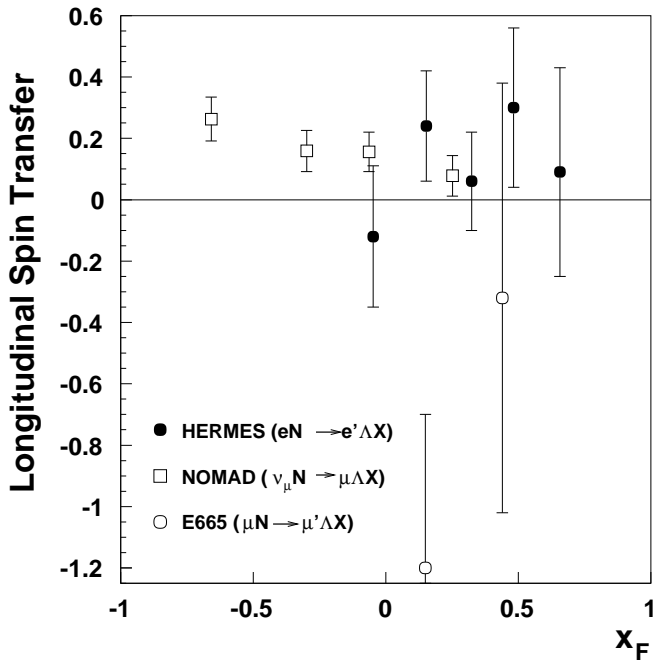}
    \caption{
      Dependence of the longitudinal spin-transfer coefficient \dll
      on $x_F$. The HERMES measurements are represented by the
      solid circles, while the open symbols represent data from
      NOMAD~\cite{NOMAD} (squares) and
      E665~\cite{E665} (circles). Error bars are statistical only.
      As explained in the text,
      for the neutrino-induced NOMAD data the quantity plotted
      is $-P^\nu_\Lambda$.
    }
    \label{fig:dll_xF}
  \end{center}
\end{figure}

%..........................................................................
The HERMES results  as a function of
$x_F$  are presented in  Fig.~\ref{fig:dll_xF} and
Table~\ref{table:dll_xF}.
%..........................................................................
% Table 2: z-dependence of DLL
\begin{table}
  \begin{center}
  \begin{tabular}{c|c||c|c}
    \hline
    $z$-range & $\langle z \rangle$ & 
	\dllnospace, axis 1 & \dllnospace, axis 2 \\
    \hline
    \hline
    $0.05 < z < 0.34$
    & $0.28$ % & $0.281$
    & $\phantom{-}0.09 \pm 0.19$
    & $\phantom{-}0.06 \pm 0.20$ \\
    $0.34 < z < 0.44$
    & $0.39$ % & $0.391$
    & $\phantom{-}0.19 \pm 0.19$
    & $\phantom{-}0.21 \pm 0.19$ \\
    $0.44 < z < 0.55$
    & $0.49$ % & $0.491$
    & $\phantom{-}0.12 \pm 0.20$
    & $\phantom{-}0.13 \pm 0.20$ \\
    $0.55 < z < 1\phantom{.00}$
    & $0.66$ % & $0.661$
    & $\phantom{-}0.03 \pm 0.23$
    & $-0.02 \pm 0.23$ \\
    \hline
  \end{tabular}
  \caption{Measured values of \dll in bins of $z$,
    for  $x_F > 0$.  The quoted uncertainties
    are  statistical only. The systematic uncertainties are on the level
    of $\pm 0.03$, as  discussed in the text.
  }
  \label{table:dll_z}
  \end{center}
\end{table}
%..........................................................................

In order  to provide a comparison with other deep-inelastic scattering
experiments, and to   illustrate the level of  agreement in the
region of overlap, the HERMES data are  shown
in Fig.~\ref{fig:dll_xF}  together with data obtained
by the NOMAD experiment~\cite{NOMAD}  at CERN
with a 43 GeV $\nu_\mu$-beam. Results from
the Fermilab E665  experiment~\cite{E665} obtained
with  a 470 GeV  polarized muon beam are also shown.

The most precise data are the charged-current
$\nu_\mu N \rightarrow \mu \Lambda^0 X$
measurements from NOMAD. The energy of the NOMAD neutrino beam
is  similar to the 27.6 GeV positron beam of the
HERMES experiment.
However, the spin-transfer coefficient \dll
presented by HERMES cannot be  immediately compared to the
longitudinal \lam polarization measured by NOMAD.
In the framework  of the quark-parton model
the polarization for the charged-current $\nu_\mu$ interaction
may be expressed as~\cite{NOMAD}:
\begin{equation}
  P^\nu_{\Lambda}(x,y,z) = -\,\frac
    { q_d(x)\,G_{1,u}^\Lambda(z)\, - \,
      (1-y)^2\,q_{\bar u}(x)\,G_{1,\bar d}^\Lambda(z) }
    { q_d(x)\,D_{1,u}^\Lambda(z)\, + \,
      (1-y)^2\,q_{\bar u}(x)\,D_{1,\bar d}^\Lambda(z) }.
  \label{eq:NOMAD_interp}
\end{equation}
(Here, $q_d(x)$ and $q_{\bar u}(x)$ represent the number densities
for quarks and antiquarks separately, contrary to the convention used in 
the rest of the paper.)
The quantity  $-P^\nu_{\Lambda}$    thus
represents the spin-transfer \dll from a struck quark to a \lam hyperon, but
for a different mixture of quark flavors than in
deep-inelastic scattering  with electron or
muon beams. However, as the NOMAD measurements were found to be nearly
independent of the variable $y$, the  interaction with $d$ quarks (which
converts $d$ quarks to $u$ quarks)  apparently
dominates over the interaction with  $\ubar$ quarks. Hence,
${-P^\nu_{\Lambda}\approx G_{1,u}^\Lambda /D_{1,u}^\Lambda}$, i.e., the
NOMAD result approximately measures the spin transfer  from $u$ quarks
to \lam  hyperons. As \lam production at HERMES is
also dominated by $u$
quark fragmentation, ${-P^\nu_{\Lambda}}$ from NOMAD can be qualitatively
compared to \dll from HERMES.  As shown in Fig.~\ref{fig:dll_xF},
the NOMAD and HERMES results
are indeed compatible in the kinematic region of overlap, $-0.2 < x_F <
0.3$. For  $x_F>0$ with average $\langle x_F \rangle=0.21$
NOMAD has obtained
${-P^\nu_{\Lambda}=0.09\pm0.06\ \mathrm{(stat)}\pm0.03\ \mathrm{(syst)}}$
which is in
very good  agreement with the  HERMES spin transfer for $x_F>0$
averaged over  the  kinematics of the experiment
($\langle x_F \rangle=0.31\pm0.01$):
${\dll = 0.11 \pm 0.10\ \mathrm{(stat)} \pm 0.03\ \mathrm{(syst)}}$.

All theoretical investigations agree that the \lam production mechanisms
for $x_F>0$ and $x_F<0$ are  different in nature.
For $x_F<0$  the average NOMAD result is
${-P^\nu_{\Lambda}= 0.21\pm0.04\ \mathrm{(stat)}\pm0.03\ \mathrm{(syst)}}$,
thus showing  a trend towards higher  positive values at negative $x_F$.
This behavior might suggest a change in the dominant  reaction
mechanism for \lam production between the current and
target-fragmentation regions, as discussed in  Ref.~\cite{Ellis}.

%..........................................................................
% Figure 5: xF-dependence of DLL + NOMAD + E665
%..........................................................................

%..........................................................................
% Table 3: xF-dependence of DLL
\begin{table}
  \begin{center}
  \begin{tabular}{c|c||c|c}
    \hline
    $x_F$-range & $\langle x_F \rangle$ & 
	\dllnospace, axis 1 & \dllnospace, axis 2 \\
    \hline
    \hline
    $ -0.2 < x_F < 0.06$        % $ -0.2 < x_F < 0.058 $
    & $-0.05$           % & $-0.047$
    & $-0.12 \pm 0.23$
    & $-0.16 \pm 0.23 $ \\
    $0.06 < x_F < 0.24$         % $0.058 < x_F < 0.238 $
    & $\phantom{-}0.15$         % & $0.153$
    & $\phantom{-}0.24 \pm 0.18$
    & $\phantom{-}0.24 \pm 0.18 $ \\
    $0.24 < x_F < 0.42$         % $0.238 < x_F < 0.421 $
    & $\phantom{-}0.32$         % & $0.324$
    & $\phantom{-}0.06 \pm 0.16$
    & $\phantom{-}0.08 \pm 0.16 $ \\
    $0.42 < x_F < 0.56$         % $0.421 < x_F < 0.555 $
    & $\phantom{-}0.48$         % & $0.483$
    & $\phantom{-}0.30 \pm 0.26$
    & $\phantom{-}0.33 \pm 0.26 $ \\
    $0.56 < x_F < 1\phantom{.00} $  % $0.555 < x_F < 1\phantom{.000} $
    & $\phantom{-}0.66$         % & $0.657$
    & $\phantom{-}0.09 \pm 0.34$
    & $-0.13 \pm 0.34 $ \\
    \hline
  \end{tabular}
  \caption{Measured values of \dll in bins of $x_F$. The quoted
    uncertainties are statistical only. The systematic uncertainties are  on
    the level of $\pm 0.03$, as discussed in the text.
  }
  \label{table:dll_xF}
  \end{center}
\end{table}
%..........................................................................
%Discussion
%...........................................................................
\subsection{D. Discussion}

The observation of a  small value of   \dll  points to
the dominance of scattering from $u$ or $d$
quarks whose polarization within the
\lam hyperons is expected to be small, although
the condition ${\Delta q^\Lambda_u=\Delta q^\Lambda_d
\equiv 0}$ does not necessarily mean that the spin-transfer
coefficient   vanishes.
Further, according to  estimates in the framework of the Lund-based
Monte Carlo model, the fraction of \lams  produced via
heavier hyperon decays  is  significant, and complicates
the production process:
only about 40\% of the \lam hyperons are produced directly from
string  fragmentation
within the $z < 0.7$ kinematic range which dominates the statistics
of the present measurement.
The \lam hyperons produced via
\sigz, \sigstar  or  \cascade decay (no other
hyperons were found to contribute significantly)
may  be polarized if their hyperon parents were produced  polarized.
For example, the average polarization of the \lam produced in the
$\sigz \rightarrow \lam \gamma$ decay is
$P_\Lambda= -\frac{1}{3}P_{\sigz}$~ \cite{Gatto}.
Since the $u$ quark is strongly polarized in the \sigz hyperon,
a nonzero spin transfer
\begin{equation}
\dll (\lam\ \textrm{from decay of}\ \sigz) =
-\frac{1}{3} \frac{\Delta q^{\sigz}_u}{q^{\sigz}_u}
\end{equation}
is  expected for  this partial channel \cite{Gustafson,Ashery}.
As the spin structures of the various hyperons differ dramatically
(e.g. in the Constituent Quark Model, 
$\Delta q^{\sigz}_u=+\frac{2}{3}$ while $\Delta q^{\lam}_u=0$),
the contributions from heavy-hyperon decay serve to dilute any net spin
transfer from the polarized struck quark to the observed \lamnospace.

In addition,  at the  moderate  beam energy of the HERMES
experiment, a contribution from
the target-fragmentation mechanism to   \lam production is not
excluded by   the requirement   $x_F>0$.
For some fraction of the events, the target-remnant diquark will be in a
spin-1 triplet state. It will be polarized since  its spin orientation is
fixed by that   of the struck quark. Hyperons produced
due to  fragmentation of the polarized diquark can
therefore also    be  polarized, 
further diluting any net spin transfer to the \lamnospace.

%==========================================================================
% Conclusions
%==========================================================================

\section{V. Conclusions}
\label{section:conclusions}

The polarization transfer from a polarized beam positron  to a
semi-inclusively
produced \lam hyperon has been studied in deep-inelastic positron
scattering at the HERMES experiment.
In the forward-production region $x_F > 0$,
and averaged over the kinematics of
the measured sample with ${\langle z \rangle =0.45 }$
and ${\langle x_F  \rangle =0.31}$,
a  spin-transfer  coefficient
${\dll = 0.11 \pm 0.10\ \mathrm{(stat)} \pm 0.03\
\mathrm{(syst)}}$ was obtained. This value  is in   good agreement
with the NOMAD result ${-P^\nu_{\Lambda} =0.09 \pm 0.06\
\mathrm{(stat)} \pm 0.03\ \mathrm{(syst)}}$ measured for ${x_F > 0}$  
(${\langle x_F \rangle =0.21}$).

The HERMES data presented here are the most precise measurements to
date of spin transfer in deep-inelastic scattering
at large $x_F$. The finding of a spin-transfer coefficient
consistent with zero is in marked contrast with the large \lam polarization
observed in $e^+ e^-$ annihilation at OPAL and ALEPH. This difference is
not unexpected, as \lam production in the reaction $e^+ e^- \ra Z^0 \ra \lam X$
is dominated by the fragmentation of strange quarks, while in
deep-inelastic scattering  the
fragmentation of $u$ and $d$ quarks,
weakly polarized in the \lam hyperon,
is the principal source.

The small spin transfer
may also be observed in   deep-inelastic   scattering
at moderate values of $z$
because of dilution of \dll due to
a significant fraction of \lam hyperons being produced via unpolarized
quarks. Although in the kinematic domain  explored experimentally
no significant   dependence of the spin transfer on either $z$
or $x_F$ is  observed, the moderate rise of \dll
at high $z$    predicted  in   Refs.~\cite{Boros,Ma02}
is not excluded by the present data.

%==========================================================================
% Acknowledgements
%==========================================================================

\begin{acknowledgments}

We gratefully acknowledge the DESY management for its support and the
staff
at DESY and the collaborating institutions for their significant effort.
This work was supported by the FWO-Flanders, Belgium;
the Natural Sciences and Engineering Research Council of Canada;
the National Natural Science Foundation of China;
the Alexander von Humboldt Stiftung;
the German Bundesministerium f\"ur Bildung und Forschung (BMBF);
the Deutsche Forschungsgemeinschaft (DFG);
the Italian Istituto Nazionale di Fisica Nucleare (INFN);
the MEXT, JSPS, and COE21 of Japan;
the Dutch Foundation for Fundamenteel Onderzoek der Materie (FOM);
the U. K. Engineering and Physical Sciences Research Council, the
Particle Physics and Astronomy Research Council and the
Scottish Universities Physics Alliance;
the U. S. Department of Energy (DOE) and the National Science Foundation
(NSF);
the Russian Academy of Science and the Russian Federal Agency for
Science and Innovations
and the Ministry of Trade and Economical Development and the Ministry
of Education and Science of Armenia.

\end{acknowledgments}

%==========================================================================
% Bibliography
%==========================================================================

% already defined by elsart
\newcommand{\etal}{\textit{et~al.} }

% Argument order: journal name, volume (boldfaced), year, page
\newcommand{\JOURNAL}[4]{#1 \textbf{#2}, #4 (#3)}

% Argument order: volume number, year, page
\newcommand{\ZPC}[3]{\JOURNAL{Zeit. Phys. C}{#1}{#2}{#3}}
\newcommand{\PRL}[3]{\JOURNAL{Phys. Rev. Lett.}{#1}{#2}{#3}}
\newcommand{\PRD}[3]{\JOURNAL{Phys. Rev. D}{#1}{#2}{#3}}
\newcommand{\PLB}[3]{\JOURNAL{Phys. Lett. B}{#1}{#2}{#3}}
\newcommand{\NPB}[3]{\JOURNAL{Nucl. Phys. B}{#1}{#2}{#3}}
\newcommand{\EPJC}[3]{\JOURNAL{Eur. Phys. J. C}{#1}{#2}{#3}}
\newcommand{\NIMA}[3]{\JOURNAL{Nucl. Instrum. Methods A}{#1}{#2}{#3}}

%========================================================================
%-------------------------
%============================================================

\end{document}